\documentclass[%
    aps,superscriptaddress,  % options for appearance
    pra, % note that prl style doesn't allow appendix and removes section numbering
    nofootinbib, %
    reprint,%twocolumn
    a4paper,% need to specify a4paper or letterpaper or watermark is messed up w/ revtex4
    longbibliography
    ]{revtex4-1} %

    \newif\ifarxiv % Comment the next line for PRL-style version
    \arxivtrue
    
\usepackage{nicefrac}
\usepackage[english]{babel}
\usepackage{soul} % strikethrough text with \sta

\usepackage{array} % for m-columns in tabular

\usepackage{hyperref}
\usepackage{xcolor}
\definecolor{mylinkcolor}{rgb}{0,0,0.5} % set link color here as red,green,blue.
\hypersetup{unicode=true, %
  bookmarksnumbered=false,bookmarksopen=false,bookmarksopenlevel=1, %
  breaklinks=true,pdfborder={0 0 0},colorlinks=true}%
\hypersetup{%
  anchorcolor=mylinkcolor,citecolor=mylinkcolor, %
  filecolor=mylinkcolor,linkcolor=mylinkcolor, %
  menucolor=mylinkcolor,runcolor=mylinkcolor, %
  urlcolor=mylinkcolor}%
\usepackage{bbm} 
\usepackage{amsmath}
\usepackage{graphicx}
\usepackage{amsthm}
\usepackage{enumerate}
\usepackage{tikz}
\usepackage{amsfonts}

\AtBeginDocument{%
    \newwrite\bibnotes
    \def\bibnotesext{Notes.bib}
    \immediate\openout\bibnotes=\jobname\bibnotesext
    \immediate\write\bibnotes{@CONTROL{REVTEX41Control}}
    \immediate\write\bibnotes{@CONTROL{%
    apsrev41Control,author="08",editor="1",pages="1",title="0",year="0"}}
     \if@filesw
     \immediate\write\@auxout{\string\citation{apsrev41Control}}%
    \fi
  }%

\makeatletter
\def\pdfstartlink@attr{attr{/Border[0 0 0 [1 5] ]/H/I/C[0 1 1]}}%
\def\@@Doi#1{\textcolor{mylinkcolor}{#1}\@@endlink}
\makeatother
\def\ForTexCount\section#1{} % trick for subcounts in TexCount, without actually adding sections.

\newcommand{\ket}[1]{\left| #1 \right\rangle}
\newcommand{\bra}[1]{\left\langle #1 \right|}

\begin{document}

\title{Self-testing of physical theories, or, is quantum theory optimal with respect to some information-processing task?}
\author{Mirjam Weilenmann}
\email{mirjam.weilenmann@oeaw.ac.at}
\affiliation{Institute for Quantum Optics and Quantum Information (IQOQI) Vienna,
Austrian Academy of Sciences, Boltzmanngasse 3, 1090 Vienna, Austria}
%\affiliation{Department of Mathematics, University of York, Heslington, York, YO10 5DD, United Kingdom}
\author{Roger Colbeck}
\email{roger.colbeck@york.ac.uk}
\affiliation{Department of Mathematics, University of York, Heslington, York, YO10 5DD, United Kingdom}
%\date{\today}
\date{$14^{\text{th}}$ January 2024}

\begin{abstract} 
  Self-testing usually refers to the task of taking a given set of observed correlations that are assumed to arise via a process that is accurately described by quantum theory, and trying to infer the quantum state and measurements. In other words it is concerned with the question of whether we can tell what quantum black-box devices are doing by looking only at their input-output behaviour and is known to be possible in several cases.  Here we introduce a more general question: is it possible to self-test a theory, and, in particular, quantum theory? More precisely, we ask whether within a particular causal structure there are tasks that can only be performed in theories that have the same correlations as quantum mechanics in any scenario. We present a candidate task for such a correlation self-test and analyse it in a range of generalised probabilistic theories (GPTs), showing that none of these perform better than quantum theory.  A generalisation of our results showing that all non-quantum GPTs are strictly inferior to quantum mechanics for this task would point to a new way to axiomatise quantum theory, and enable an experimental test that simultaneously rules out such GPTs.
\end{abstract}

\maketitle

\ifarxiv\section{Introduction}\else\noindent\textit{Introduction.}|\fi When we do experimental science we are limited to classical data in the sense that the measurement settings we choose and the outcomes we observe are always classical. However, we can use quantum systems to model how the classical outputs come about.  For example, quantum theory might tell us that when a particular (classical) setting is chosen, this corresponds to the preparation of some quantum state $\ket{\psi}$. Similarly, another (classical) setting may correspond to a particular quantum measurement (say the Positive-Operator-Valued-Measure (POVM) $\{E_x\}_x$), and according to quantum theory the Born rule gives us the outcome probabilities $\bra{\psi}E_x\ket{\psi}$. Often we want to go the other way.  Suppose, for instance, that we wish to infer the quantum description of the state generated in a particular process. Given sufficiently many trusted measurement devices we may be able to do this using state tomography~\cite{Fano}.

In some cases, trusted devices are not needed. For these, we can infer the quantum state and measurements (up to some symmetries) based only on the observed correlations of classical variables.  This is referred to as \emph{self-testing}~\cite{MayersYao}. For instance, by observing correlations that saturate Tsirelson's bound~\cite{Cirelson}, i.e., that reach the maximal score allowed by quantum mechanics in the CHSH game~\cite{CHSH}, one can infer that the quantum state must be a singlet up to local isometries~\cite{PopescuRohrlich}. Such work was extended to GHZ states~\cite{GHZ} in~\cite{ColbeckThesis,CK2} and subsequent work has shown that all two-qubit pure entangled states can be self-tested~\cite{CGS17}.

Self-testing as introduced above is thus the question of whether a quantum description of some devices can be inferred by only interacting with them in a classical way.  Significantly, it assumes that there is a valid quantum description.  In the present work we go beyond this to consider the question of what happens when this assumption is dropped.  Can we, by only relying on an observed classical input-output behaviour, infer that a particular theory must be used to generate it?  This is the question of self-testing of a physical theory. Of most interest to us will be self-testing of quantum theory.  If we can find a task within a particular causal structure\footnote{In the context of the present work these are diagrams indicating the allowed flow of information during the game. For a more technical account see, e.g.,~\cite{review}.} that cannot be performed in any theory other than quantum mechanics, then we would have a direct test of quantum mechanics.  This could also constitute an insight into the question of ``why quantum theory?''\bigskip

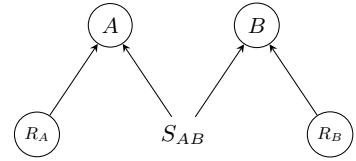
\begin{figure}
\centering
\resizebox{0.75\columnwidth}{!}{%
\begin{tikzpicture} [scale=1.0]
\node[draw=black,circle,scale=0.75] (B1) at (4,-0.5) {$R_A$};
\node[scale=1] (B2) at (6,-0.5) {$S_{AB}$};
\node[draw=black,circle,scale=0.75] (B3) at (8,-0.5) {$R_B$};
\node[draw=black,circle,scale=0.95] (B4) at (5,1) {$A$};
\node[draw=black,circle,scale=0.95] (B5) at (7,1) {$B$};
\node (0) at (9.2,1) {};
\node (00) at (2.8,1) {};

\draw [->,>=stealth] (B1)--(B4);
\draw [->,>=stealth] (B2)--(B4); %node [below,pos=0.8,yshift=-1.5ex] {$S_A$};
\draw [->,>=stealth] (B2)--(B5); %node [below,pos=0.8,yshift=-1.5ex] {$S_B$};
\draw [->,>=stealth] (B3)--(B5);  
\end{tikzpicture}
}
\caption{The bipartite Bell causal structure and CHSH game. A referee asks question $R_A$ to Alice and $R_B$ to Bob, choosing these uniformly. Their respective answers to these questions are $A$ and $B$. These may be formed by measurement on part of a shared bipartite state $S_{AB}$, whose most general form depends on the theory under consideration (in quantum theory it can be any shared quantum state, for instance). In the case where $A$, $B$, $R_A$ and $R_B$ are binary, the CHSH game is said to be won if $A\oplus B=R_A\cdot R_B$, where $\oplus$ denotes addition modulo $2$. Note that in this causal structure the measurements only act on single subsystems of the shared state.}
\label{fig:Bell}
\end{figure}

\ifarxiv\section{Self-testing of physical theories} In this paper \else\noindent\textit{Self-testing of physical theories.}|In this Letter \fi we discuss self-testing of physical theories and a candidate task for doing so for quantum theory.  Since dealing with arbitrary alternatives to quantum theory is challenging we begin the investigation by restricting our considerations to generalised probabilistic theories (GPTs)~\cite{Hardy2001, Barrett2007} (i.e., \emph{GPT self-testing}). These are a class of theories that include classical and quantum theory as well as theories with more non-local correlations still. In this framework, the state space of a system can be expressed as a compact convex set in a real vector space and measurements are collections of effects, linear maps from states to probabilities. Joint states can be defined for multipartite systems, and we require that these lead to non-signalling correlations.\footnote{While non-signalling is essential so that considering causal structures makes sense, assuming local tomography is not necessary here.} In this class, there is a well-known theory that is maximally non-local with respect to bipartite correlations, referred to as \emph{box-world}, in which PR-box correlations (i.e., those that can win the CHSH game described in Figure~\ref{fig:Bell} with probability $1$) are realisable.  

In the present work we consider \emph{correlation self-testing} in which we seek a causal structure and a task such that any GPT that achieves the (optimal) performance in that task must have the same set of possible correlations as quantum theory in any causal structure. There are several other forms of self-testing one could consider, aiming to identify, for instance, the state space of the theory, or, more generally, the states, measurements and possible transformations. We discuss these notions in our accompanying paper~\cite{selftest_PRA}.  

Given two theories $\mathcal{T}_1$ and $\mathcal{T}_2$, if there is a causal structure in which $\mathcal{T}_1$ generates some correlations that cannot be generated in $\mathcal{T}_2$, then there is a task in which $\mathcal{T}_1$ outperforms $\mathcal{T}_2$. Similarly, suppose we have a family of theories $\left\{\mathcal{T}_i \right\}_{i=1}^n$ in which for each pair there is some causal structure for which $\mathcal{T}_i$ generates some correlations that cannot be generated in $\mathcal{T}_j$. Then $n-1$ tasks are sufficient for correlation self-testing any theory within this set in the sense that by requiring a certain performance in each task, each theory can be singled out uniquely.  In general we may want to consider infinite sets of theories. Correlation self-testing of quantum theory then boils down to identifying a family of tasks|optimally a single one|that are sufficient for singling it out. In this Letter, we propose a task, the adaptive CHSH game, and explain why it is a promising candidate for self-testing quantum theory.

We also note that if it is the case that quantum theory leads to a subset of the correlations of another theory, $\mathcal{T}$, in \emph{all} causal structures, and where this inclusion is strict in some of them, then quantum theory cannot be fully correlation self-tested.  Instead, our inability to find non-quantum correlations would make $\mathcal{T}$ less plausible. Furthermore, designing a self-test for $\mathcal{T}$ would lead to insights about quantum theory and point to possible additional axioms that might rule $\mathcal{T}$ out.

Before introducing the adaptive CHSH game, we make a few simple observations regarding the self-testing of other theories.  Were we to observe PR-box correlations in the bipartite Bell causal structure (see Figure~\ref{fig:Bell}) we would be able to self-test the set of all non-signalling correlations with two possible inputs and outputs per party. This would correspond to a large restriction on the underlying theories to those in which PR-box correlations can be distilled. 

Conversely, finding correlations achieving the quantum bound in the CHSH game eliminates theories for which these cannot occur, but does not rule out more non-local theories (e.g., those in which PR-boxes could be realised).  This observation is somewhat trivial.  However, it is useful for illustrating an important point: correlation self-testing of quantum theory cannot be done in any causal structure in which all the measurements can only act on single subsystems.  In such causal structures, moving to a GPT that has a larger set of states cannot decrease the set of realisable correlations and hence quantum theory cannot be optimal for any task in such a causal structure (or, more precisely, if it is, it shares this optimality with all theories with larger sets of correlations).

GPTs that have larger state spaces than quantum theory have smaller effect spaces. This is because a given effect must be applicable to any of the allowed states, i.e., it must map them to a probability. Adding to the state space corresponds to additional requirements for an effect to be valid\footnote{In a GPT, the set of allowed effects must be a subset of the dual cone to the cone of subnormalised states, so that, without further restrictions large state spaces have small effect spaces and vice versa.}.  In order to find a causal structure in which quantum theory allows correlations that cannot occur within any other GPT we need to simultaneously exploit the need for a sufficiently large state space and a sufficiently large effect space.

\bigskip

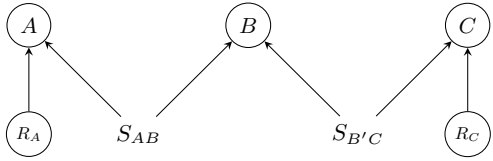
\begin{figure}
 \centering
\resizebox{0.75\columnwidth}{!}{%
 \begin{tikzpicture} [scale=1.5]
 \node[draw=black,circle,scale=0.95] (1) at (0,1) {$A$};
 \node[draw=black,circle,scale=0.75] (R1) at (0,0) {$R_A$};
 \node (2) at (1,0) {$S_{A B}$};
 \node[draw=black,circle,scale=0.95] (4) at (2,1) {$B$};
 \node (5) at (3,0) {$S_{B' C}$};
 \node[draw=black,circle,scale=0.95] (6) at (4,1) {$C$};
 \node[draw=black,circle,scale=0.75] (R2) at (4,0) {$R_C$};
 \draw [->,>=stealth] (R1)--(1);
 \draw [->,>=stealth] (2)--(1);
 \draw [->,>=stealth] (2)--(4);
 \draw [->,>=stealth] (5)--(4);
 \draw [->,>=stealth] (5)--(6);
 \draw [->,>=stealth] (R2)--(6);
\end{tikzpicture}
}
\caption{Causal structure of the adaptive CHSH game. The values of $A$, $B$ and $C$ as well as the referee's questions $R_A$ and $R_C$ determine whether the game is won. Resources are shared between $A$ and $B$, and between $B$ and $C$, but there are no shared tripartite resources.  Although $A$ and $C$ can be formed by measurements on single subsystems, the value of $B$ can be obtained by jointly measuring the $B$ part of $S_{AB}$ and $B'$ part of $S_{B'C}$.}
\label{fig:game}
\end{figure}

\ifarxiv\section{Correlation self-testing of quantum theory in the adaptive CHSH game} \else\noindent\textit{Correlation self-testing of quantum theory in the adaptive CHSH game.}|\fi To pursue this we consider the causal structure of Figure~\ref{fig:game} and a game played by three cooperating players, Alice, Bob and Charlie, which we call the adaptive CHSH game. A referee asks Bob to choose one of four versions of the CHSH game~\cite{CHSH}. We call this choice $B$. The referee then asks Alice and Charlie questions, denoted $R_A$ and $R_C$ respectively, for which they have to give answers, labelled $A$ and $C$ respectively, where $R_A$, $R_C$, $A$ and $C$ can take values $0$ or $1$, and $R_A$ and $R_C$ are chosen uniformly at random by the referee. The three players win the game if Alice and Charlie's answers win the instance of the CHSH game Bob chose\footnote{Bob's choice is not communicated to Alice and Charlie so that they are oblivious to which game they are playing (they may even give their answers before Bob).}.

The instances of the CHSH game are as follows:
\begin{center} \begin{tabular}{| c | c |} 
\hline
$B$ (Bob's choice) & Winning condition\\
\hline
$(0,0)$ & $(r_A\oplus1)\cdot r_C=a\oplus c$ \\
\hline
$(0,1)$ & $(r_A\oplus1)\cdot(r_C\oplus1)\oplus1=a\oplus c$ \\
\hline
$(1,0)$ & $(r_A\oplus1)\cdot(r_C\oplus1)=a\oplus c$ \\
\hline
$(1,1)$ & $(r_A\oplus1)\cdot r_C\oplus1=a\oplus c$ \\
\hline
\end{tabular}
\end{center}
Thus, the overall winning probability of the game for a strategy that leads to a distribution ${P_{ABCR_AR_C}(a,b,c,r_A,r_C)}$ is 
\begin{align*}
p_\mathrm{win}=\!\!\!\!\!\!\sum_{a,b,c,r_A,r_C}\!\!\!\!\!\!P_{ABCR_AR_C}(a,b,c,r_A,r_C) Q(a,b,c,r_A,r_C),
\end{align*} 
where $Q(a,b,c,r_A,r_C)$ is $1$ if the corresponding winning condition in the above table is met and $0$ otherwise.  Since $R_A$ and $R_C$ are chosen uniformly, we have $P_{ABCR_AR_C}(a,b,c,r_A,r_C)=\frac{1}{4}P_{ABC|r_Ar_C}(a,b,c)$.

The idea behind the use of this game is that in order to win with the highest possible probability in quantum mechanics, an entanglement swapping operation is required. If $S_{AB}$ and $S_{B'C}$ are both maximally entangled qubit pairs, and a Bell basis measurement is performed on the $BB'$ systems to give the outcome $B$, then, by choosing the measurements generating $A$ and $C$ appropriately, the game can be won with probability $\frac{1}{2}\left(1+\frac{1}{\sqrt{2}}\right)$ (see our accompanying paper~\cite{selftest_PRA} for an explicit strategy).  In order that a value as high as this can be obtained it is necessary that the same value is obtainable in the CHSH game in the bipartite Bell causal structure of Figure~\ref{fig:Bell}.  Therefore, in any GPT for which the maximum probability of winning the CHSH game is below Tsirelson's bound of $\frac{1}{2}\left(1+\frac{1}{\sqrt{2}}\right)$~\cite{Cirelson}, the same upper bound holds for the adaptive CHSH game.  Quantum theory hence beats any theory for which Tsirelson's bound cannot be saturated.

More interestingly, in all GPTs whose joint state space is formed by taking the maximal tensor product of the single-system state space, the joint measurement on $BB'$ can be thought of in terms of a measurement on one of the subsystems, followed by a measurement on the other (possibly depending on the result), or convex mixtures of measurements of this kind~\cite{Barrett2007}.  Such measurements cannot lead to any resulting non-locality between $A$ and $C$, even after conditioning on the outcome, and hence in such theories the adaptive CHSH game cannot be won with probability greater than $3/4$ (the maximum achievable classically). An example of such a theory is box-world, see~\cite{Barrett2007,GMCD}. For this particular theory, previous work on \emph{couplers} also directly implies this restriction~\cite{Short2006, Skrzypczyk2009b, Skrzypczyk2009}.

From these arguments, when considering the adaptive CHSH game, \ifarxiv\else it can be shown that \fi quantum theory is superior to any theories in which the joint states are formed with either the minimal tensor product (where all states are separable) or the maximal tensor product. These behave like two extremes\footnote{For GPTs that are locally tomographic these tensor products are known to define the minimal and the maximal joint state spaces that are possible.}, with the tensor product of quantum theory sitting in between. The adaptive CHSH game is hence a reasonable candidate for a task for which quantum theory is optimal among all GPTs. 

In order to show that quantum theory cannot be beaten, we need to consider GPTs whose joint state space is formed using other tensor products. However, with the exception of quantum theory\footnote{There were indications that the same is true for quantum theory over the real numbers~\cite{MMG}, but it has since been shown that there is a separation~\cite{RealQM}.}, we are not aware of explicit alternative tensor products. Nonetheless, the single system state spaces already impose restrictions on the achievable correlations that allow us to bound their performance in the adaptive CHSH game.

To do so we consider a set of self-dual theories in which the single system state spaces take the form of a two-dimensional regular polygon with a varying number of sides, $n$~\cite{Janotta2011, NoR2}. Roughly speaking, self-dual theories are those for which the set of (subnormalized) states and effects are the same (for instance, in quantum theory, subnormalized states are positive operators with trace at most $1$, as are the POVM elements, which correspond to effects).  In the case of odd $n$, taking the effect space to be the maximal possible, i.e., the set of all linear maps from states to probabilities, naturally yields a self-dual theory, while for even $n$ one has to transform these spaces (see~\cite{NoR2} and our accompanying paper~\cite{selftest_PRA}). Our main result for these GPTs is the following.

\smallskip

{\it Theorem (informal):} Consider any GPT where the single-system state space is a self-dualized regular polygon system with up to $30$ extremal states and the joint state space is any tensor product between the minimal and the maximal. If the sources $S_{AB}$ and $S_{B'C}$ each correspond to one pair of such gbits then the winning probability in the adaptive CHSH game is upper bounded by the quantum value.

\smallskip

\begin{figure}
	\includegraphics[width=0.95\columnwidth]{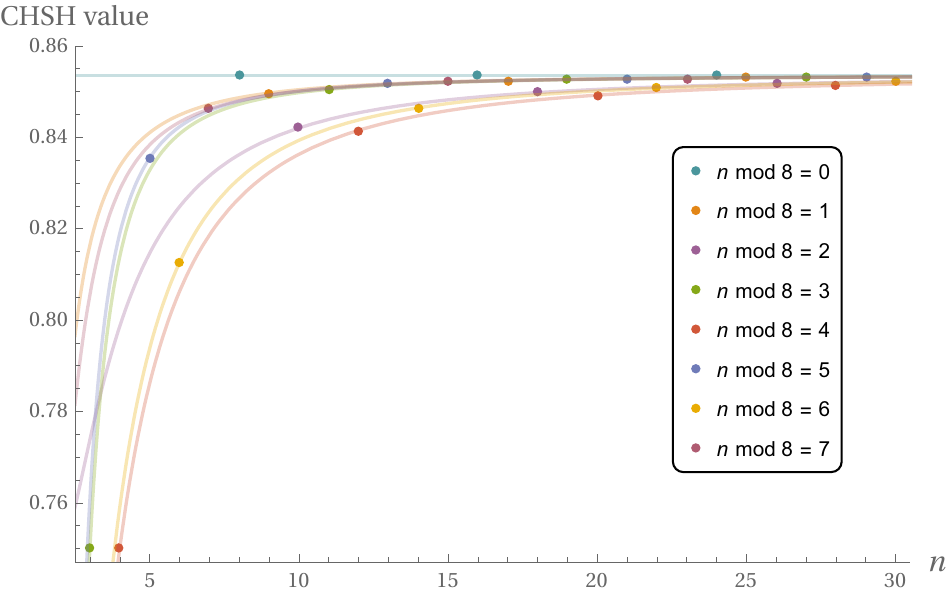}
	\caption{Maximal winning probability in the CHSH game achievable in the maximal tensor product of self-dualized polygon systems. Along the horizontal axis we display the number of extremal vertices $n$ of the local state space, while on the vertical axis we display the respective winning probability. The points are the values obtained in our optimizations while the curves depict respective formulae these values follow. The colours are used to group points that appear to follow the same analytic formula.}
	\label{fig:plot}
\end{figure}

To prove this we use the fact that the winning probability in the adaptive CHSH game can be upper bounded by the maximum probability of winning the regular CHSH game. Using linear programming we evaluated the latter using any bipartite state from the maximal tensor product of the considered state spaces (for other tensor products these are upper bounds). The values obtained are illustrated in Figure~\ref{fig:plot}, and we found analytic formulae that appear to repeat modulo 8, each of which tends to the quantum value as $n$ tends to infinity. For $n\!\mod 8\neq 0$ the values are strictly smaller than the maximal quantum value, while for $n\!\mod 8=0$ the upper bound is equal to the quantum value. However, we do not expect any of these upper bounds to be achievable in the adaptive CHSH game because they are based on the maximal tensor product in which case we know that the adaptive CHSH game cannot be won with probability larger than $3/4$.

Thus, the above theorem is expected to generalize to any $n$ and to a statement that the winning probabilities are strictly smaller than the quantum one even for $n\!\mod8=0$.  Further details of these results as well as a discussion of bipartite polygon state spaces without self-duality can be found in our accompanying paper~\cite{selftest_PRA}.

A number of challenges need to be overcome to extend our results to a full correlation self-testing of quantum theory.  Firstly, there is a wide variety of possible theories, and not many theoretical results have been developed that can simultaneously apply to them all.  Secondly, we lack general bounds on when non-locality can be distilled, and by how much. For instance it could be that individual systems in a particular GPT always give rise to correlations between $A$ and $C$ that cannot violate Tsirelson's bound.  However, if, given many copies of such a system, we could distil correlations that violate Tsirelson's bound without communication, then the theory could win the adaptive CHSH game better than quantum theory.  Thirdly, there is a lack of known ways to construct joint state spaces other than with the minimal and maximal tensor products, both of which we know do not give an advantage in this game.  Thus, the question we have asked necessitates the development of several lines of research before it can be answered in full generality. In an accompanying paper~\cite{selftest_PRA} we report additional progress in this direction.\bigskip

\ifarxiv\section{Related Work}\else\noindent\textit{Related Work.}|\fi The question ``why quantum mechanics?''\ has been a topic of debate since the conception of the theory.  
Unlike other theories that have firm physical principles behind them (such as the principle that the laws of physics are the same in every reference frame for special relativity), quantum mechanics is usually presented as a series of mathematical axioms whose underlying physical significance is unclear. There have been numerous attempts to give quantum mechanics a more physical axiomatisation, going back to Popescu and Rohrlich~\cite{Popescu1994} who asked whether quantum mechanics is the most non-local theory that obeys the no-signalling principle. That it is not follows because PR-boxes are non-signalling but can win the CHSH game of Figure~\ref{fig:Bell} with probability $1$, in violation of Tsirelson's bound~\cite{Cirelson} (the PR-box correlations and the impossibility of realising them in quantum theory was also realised by Tsirelson~\cite{Cirelson93}).

There have been several other attempts to find such a principle from which quantum theory naturally follows~\cite{Brassard2006,Linden2007,Pawlowski2009,Navascues2010a,Fritz2013b}. Each of these principles imposes a restriction on the correlations a `reasonable' theory may produce, however, none of them singles out the set of quantum correlations exactly. Instead, they are also obeyed by correlations that are not achievable quantum mechanically, in particular by the set of \emph{almost quantum correlations} (potentially with the exception of Information Causality, for which numerical evidence supports this, but it is not strictly proven)~\cite{Navascues2015}. We also remark that establishing optimality with respect to a task, as considered in this \ifarxiv paper, \else Letter, \fi is more objective than deciding whether axioms seem `reasonable'.\bigskip

\ifarxiv\section{Conclusions}\else\noindent\textit{Conclusions.}|\fi We have discussed the possibility of showing optimality of quantum theory using the adaptive CHSH game, but there are many other tasks that could be considered instead, and it could be the case that a family of tasks are required. In particular, our game only involves sharing bipartite systems, whereas consideration of more parties may be necessary to single out quantum mechanics. In addition, although we have phrased the task via a game, we could alternatively look to find a set of correlations achievable in a causal structure within quantum theory but not in any other GPT. The triangle causal structure~\cite{Fritz2012} is a candidate for this because it also has the possibility of joint measurements on multiple subsystems.  

If it is possible to show that the optimal performance in the adaptive CHSH game (or an alternative) is only achieved by theories with the same correlations as quantum theory, this will point towards a new way to axiomatise quantum theory. Admittedly, an axiom of the form ``the theory is the one with the highest winning probability with respect to the adaptive CHSH game'' would not be especially natural. However, there may be a deeper and more natural principle underlying it.  More significantly, attempting to experimentally observe the correlations that optimally win the game would be a way to directly rule out a range of alternative theories, hence providing a strong confirmation of the validity of quantum mechanics. This could have impact on the search for future theories, such as candidates for uniting quantum mechanics and gravity, although at the moment this remains speculative. Nevertheless, we consider this a promising research direction that has the potential to be transformative.\bigskip

\ifarxiv\acknowledgements\else\noindent\textit{Acknowledgements.}|\fi This work was supported by an EPSRC First Grant (grant number EP/P016588/1) and by the Austrian Science fund (FWF) stand-alone project P~30947.\bigskip

%\bibliography{hplusgame}

%

\end{document}